\documentstyle[twoside,fleqn,espcrc2]{article}

\newcommand{\AmS}{{\protect\the\textfont2
  A\kern-.1667em\lower.5ex\hbox{M}\kern-.125emS}}

\def\Journal#1#2#3#4{{#1} {\bf #2}, #3 (#4)}

\def\PRD{{\em Phys. Rev.} D}

\def\be{\begin{equation}}
\def\ee{\end{equation}}
\def\bea{\begin{eqnarray}}
\def\eea{\end{eqnarray}}
\def\sla{\raise.15ex\hbox{$/$}\kern-.57em}

\title{
\vspace{-5.0cm}
\begin{flushright}{\normalsize RUHN-99-2}\\
\end{flushright}
\vspace*{2.5cm}
Tricks to implement the overlap Dirac operator
\thanks{Talk at Lattice '99, June 29 - July 3, Pisa, Italy.}
}
\author{Herbert Neuberger\address{Department of Physics and Astronomy, \\ 
        Rutgers University, Piscataway, NJ 08855-0849}%
        \thanks{Research supported in part by the DOE, 
grant \#DE-FG05-96ER40559.}
        }
\begin{document}

\begin{abstract} 

I present several tricks to help implement the overlap Dirac operator
numerically. 
\end{abstract}
\maketitle

\section{Introduction}
There are new ways to implement chirality exactly on the lattice. This
theoretical progress can be implemented numerically in a myriad of
ways. I am not sure that it makes sense to have all the larger machines,
(QCDSP Columbia and Riken/BNL, CP-PACS) do domain walls of exactly the
same type. After all, this is only one of many possible truncations of
the overlap. The SCRI and Kentucky groups have been more daring and 
innovative and, I think, their results show that it paid off.
My purpose in this talk is to present a few variations on the topic
of direct numerical implementation of the overlap Dirac operator $D_0$.

The plan is to first present the basic procedure \cite{prl} and then
proceed to describe five ``tricks''.

\section{Basics and refinements}
\subsection{Basic procedure}
The objective is: Given a $\psi$ compute $\chi=D_o\psi\equiv 
\left ( 1+\gamma_5 \varepsilon (H_W )\right ) \psi$. The basic method uses a rational 
approximation to the sign function \cite{higham} 
\begin{eqnarray*}
&\varepsilon_n (x) \equiv \varepsilon^{(1)}_n (x)  = {x\over n}
\sum_{s=1}^n {1\over {x^2 \cos^2 \theta_s + \sin^2 \theta_s }} \equiv \\
&{{(1+x)^{2n} - (1-x)^{2n}}\over {(1+x)^{2n} + (1-x)^{2n}}}\equiv
\tanh (2n \tanh^{-1} (x)) \equiv \\ &\tanh (2n\tanh^{-1} ({1\over x}),\\
\end{eqnarray*}
where $\theta_s = {\pi\over{2n}} (s- {1\over 2 })$.
Numerically the main point is that using the SESAM shifted mass trick the
cost of computing $\sum_{s=1}^n {1\over {x^2 \cos^2 \theta_s + 
\sin^2 \theta_s }}\psi$ in floating operations is roughly the same as the cost
of the single inversion  ${1\over {x^2 \cos^2 {\pi\over{4n}} + 
\sin^2 {\pi\over{4n}} }}\psi$.
For the inversion we use the conjugate gradient (CG) algorithm. Memory
usage grows linearly with $n$. 

$\varepsilon^{(1)}_n (x)$ has some important properties:
\begin{eqnarray*}
&\varepsilon^{(1)}_n (x)=-\varepsilon^{(1)}_n (-x)=
\varepsilon^{(1)}_n ({1\over x})\\
&|\varepsilon^{(1)}_n (x)|\leq
\varepsilon^{(1)}_n (\pm 1)\equiv 1.\\
\end{eqnarray*}
Pick an $n$ such that 
$\varepsilon^{(1)}_n (x)\approx \varepsilon (x)$
for $x\in [-A,-{1\over A}] \cup [{1\over A}, A ],~A\geq 1$ and pick
$x=\lambda H_W$, with the spectrum of $|H_{W}|$ bracketed between
$\lambda_{\rm min}$ and $\lambda_{\rm max}$. Choose $\lambda$ so that
$A=\lambda\lambda_{\rm max}={1\over{\lambda \lambda_{\rm min}}}$, that is
$\lambda={1\over{\sqrt{\lambda_{\rm min}\lambda_{\rm max}}}}$. Let 
$\kappa={\lambda_{\rm max}\over
\lambda_{\rm min}}$ be the condition number. For the approximation
to be good we need $2 n >> A = \sqrt{\kappa}$.
The problem becomes, as always, that one needs large $n$ if the condition
number is large. 
\subsection{Trick 1}
SCRI\cite{scri} used another rational approximation for the sign function,
$\varepsilon^{(2)}_n (x)$. This
approximation is optimal in the $\infty$-norm and the coefficients
of the fraction are computed using the Remez algorithm. Thus,
one achieves better accuracy with a smaller $n$. But, 
$|\varepsilon^{(2)}_n (x)|$ no longer is bounded by unity. There
is therefore the danger of producing unphysical zeros in $D_o$.
The trick I suggest is to combine and use 
$\varepsilon^{(12)}_{nm} (x) =\varepsilon^{(1)}_n (\varepsilon^{(2)}_m (x))$ 
to recover the bound.
\subsection{Trick 2}
Here I am concerned with memory usage, something that can affect
performance dramatically when cache is exceeded. The idea is to
use a two pass shifted-mass CG. This is similar to a standard
procedure applied to Lanczos diagonalization when an eigenvector
is also desired. In exact arithmetic the algorithm is the same as
the basic one. The cost in floating point operations is at most
a factor of 2, but on a RISC processor with standard (high) cache
miss penalty one finds much smaller costs for practically
interesting values of $n$ \cite{twopass}.
\subsection{Trick 3}
The main problem in implementations is that at desirable
gauge couplings one often encounters eigenvalues of $H_W$ very close to zero.
But, the physics behind the overlap construction allows to replace
the argument of the sign function with any reasonable lattice
version of the hermitian Dirac operator in the continuum with
a large negative mass term. Thus, there is no direct reason
for the argument of the sign function to often have a spectrum
extending too close to the origin. The overlap itself can provide
replacements of $H_W$, $H^\prime_W$ that are better in this respect.
The idea is to use a rough approximation to the sign function,
$\varepsilon_{\rm rough} (x)$, which is fast to implement and take
$H^\prime_W = \rho \gamma_5 + \varepsilon_{\rm rough} (H_W )~~0<\rho < 1$.
The choice for $\rho$ makes the physical Dirac mass negative and, if
$\varepsilon_{\rm rough}$ were a good approximation to the sign
function, $H^\prime_W$ would have a gap in its spectrum around the
origin of size $|\rho -1|$ and a condition number ${{|\rho +1|}\over
{|\rho -1|}}$. So, the suggestion is to plug $H^\prime_W$ into
$\varepsilon^{(12)}_{nm}$ and use either the basic procedure or its
two pass version. The distinguishing feature of this trick is that
it uses some physics input. 
\subsection{Trick 4}
The numerical difficulties are caused by the nonanalyticity of the sign 
function at zero. The idea here is two double the number of fields
so as to ameliorate the singularity in the sign function. In this 
way one may hope to avoid nested CG if one can replace the rational 
approximation by a polynomial of moderate degree. 
To see how this could possibly work, \cite{dyna},
introduce the fields $ \bar \chi = \pmatrix {\bar\psi & \bar \phi \cr}$ and 
$\chi = \pmatrix {\psi\cr\phi\cr} $. Next, 
consider the following identity, easily proven by Gaussian integration
over $\bar\phi , \phi$:
\begin{eqnarray*}
\int&  d\bar\phi d\phi  e^{\bar\chi \pmatrix {\gamma_5 & (H^2_W )^{1/4}\cr 
(H^2_W )^{1/4} & -H_W \cr} \chi} = \\
& \det H_W  e^{\bar\psi (\gamma_5 +
\varepsilon (H_W )) \psi }\\
\end{eqnarray*}
The main point is that
$|x|^{1\over 2}$ is less violently behaved at the origin than
$1/|x|$ and might be easier to reproduce either polynomially,
or by a low $n$ rational. In a dynamical simulation the $\det H_W$
prefactor will need to be canceled by pseudoferminos. The important
point is that the induced action for the $\bar\psi \psi$ fields has
the right structure. 
\subsection{Trick 5}
The moral from trick 4 is that adding extra fields to induce the
desired action for the fields $\bar\psi\psi$ softens the singularity of
$\varepsilon$. Theoretically, we know that adding and infinite number
of fields removes the singularity altogether. For an approximation to the
sign function characterized by order $2n$ one expects that the addition
of $2n$ fields can remove all polynomials or rationals altogether. This
brings the approximation closer to domain walls, but maintains a larger
degree of flexibility. 

The trick I am describing below \cite{dyna} rests on two observations: (1)
Any rational approximation can be viewed as a truncated continued fraction,
which, when untruncated, would represent the sign function exactly (except
exactly at the origin, where the sign function isn't defined) (2)
Any (truncated) continued fraction can be exactly mapped into a (finite) chain
model. Rather than presenting the idea in the abstract let
us focus on a chain realization of $\varepsilon^{(1)}_n (x)$. The general
case will become obvious.

First, the rational approximation
has to be written in the form of a continued fraction
with entries preferably linear in $H_W$. 
I start from a formula that goes as far back as Euler (see below), 
and subsequently
use the invariance under inversion of $x$ to move the $x$ factors
around, so that the entries become linear in $x$.   
\begin{eqnarray*}
&\varepsilon_n (x) =\\
&{ {\displaystyle 2nx }\over \displaystyle 1 +
                 {\strut {\displaystyle (4n^2-1)x^2 } 
\over\displaystyle 
                                      3+
                   {\strut {\displaystyle (4n^2-4)x^2 } 
\over\displaystyle 
                                      5 + \dots
                   {\strut \displaystyle   \ddots \over 
{\displaystyle 4n-3}+
                     {\strut {\displaystyle 
[4n^2 - (2n-1)^2]x^2} \over 
{\displaystyle 4n-1} }}}}}\\
\end{eqnarray*}
Now, with the help of extra fields, I write a Gaussian path integral
which induces the desired action between a chosen subset of fields:
\begin{eqnarray*}
\int d\bar\phi_1 d\phi_1  d\bar\phi_2 d\phi_2 \dots  d\bar\phi_n d\phi_n  e^{S_*} =\\
(\det H_W )^{2n} 
 e^{-\bar\psi (\gamma_5 +\varepsilon_n (H_W ) )\psi }\\
\end{eqnarray*}
The quadratic action $S_*$ couples the 
extended fermionic fields $\bar \chi , \chi$:
\begin{eqnarray*}
\bar \chi = \pmatrix {\bar\psi & \bar\phi_1 & \dots & \bar\phi_{2n}\cr},~~
\chi  =  \pmatrix {\psi & \cr\phi_1 & \cr\vdots &\cr \phi_{2n}} \\
\end{eqnarray*}
$S_* = \bar\chi {\bf H} \chi$, where the new kernel, ${\bf H}$, 
has the following block structure:
\begin{eqnarray*}
\pmatrix{-\gamma_5 & \sqrt {\alpha_0 }& 0&  \dots &\dots & 0 \cr
          \sqrt {\alpha_0 }& H_W & \sqrt {\alpha_1 }& \dots &\dots &0\cr
           0& \sqrt {\alpha_1 } & -H_W &  \dots&\dots & 0\cr
                  \dots & \dots & \dots & \ddots & \dots & 0\cr
                   \dots & \dots & \dots & \dots & H_W & 
                                            \sqrt{\alpha_{2n-1}}\cr
	          \dots & \dots & \dots & \dots & \sqrt{\alpha_{2n-1}}& -H_W \cr}\\
\end{eqnarray*}
The numerical coefficients $\alpha$ are given below:
\begin{eqnarray*}
\alpha_0 =2n ,~
\alpha_j = {{(2n-j)(2n+j)}\over {(2j-1)(2j+1)}},~ j=1,2,...\\
\end{eqnarray*}
The hope is that the condition number of ${\bf H}$ will be manageable.

So, at the expense of adding extra fields one can avoid
a nested conjugate gradient procedure when dynamical fermions are simulated.
The chain version of the direct truncation of the overlap
Dirac operator is similar in appearance to domain walls. But, one is free
to change both the rational approximation and its chain implementation. 

Moreover,
since here the argument of the approximated sign function is $H_W$, not
the rather cumbersome logarithm of the transfer matrix of the domain wall
case, eigenstates of $H_W$ with small eigenvalues  can be eliminated by
projection with greater ease \cite{scri}. This elimination, although costly
numerically, vastly increases the accuracy of the approximation to the sign
function. Actually, at this stage of the game and at practical gauge
coupling values, the use of projectors seems to be numerically indispensable
to direct implementations of the QCD overlap Dirac operator. 
But, no projectors have been implemented in domain wall
simulations. However \cite{vectlike}, the domain
wall version is too close to the overlap
Dirac operator based on $H_W$ to 
believe that projections are necessary
in one case but can be ignored in the other. 
Thus, I urge caution when interpreting data 
obtained using very light domain wall fermions.
Domain wall practitioners might consider implementing
projectors to improve their reach to low quark masses.

\section{Final comments}
Practical tests of the above tricks are both badly needed and embarrassingly
few at the moment. There is not much to test in Trick 1. Trick 2 has been
tested - its usefulness is architecture dependent. Tricks 3 through 5 have
not been tested yet. Still, I think 
it is important to share insights and maintain
flexibility, so I decided to present these ideas at an early stage.

\end{document}